# Extending the class of solvable potentials
## III. The hyperbolic single wave

H. Bahlouli[a,b] and A. D. Alhaidari[b,c*]

[a]*Physics Department, King Fahd University of Petroleum & Minerals, Dhahran 31261, Saudi Arabia*
[b]*Saudi Center for Theoretical Physics, Dhahran, Saudi Arabia*
[c]*KTeCS, P.O. Box 32741, Jeddah 21438, Saudi Arabia*

A new solvable hyperbolic single wave potential is found by expanding the regular solution of the 1D Schrödinger equation in terms of square integrable basis. The main characteristic of the basis is in supporting an infinite tridiagonal matrix representation of the wave operator. However, the eigen-energies associated with this potential cannot be obtained using traditional procedures. Hence, a new approach (the "potential parameter" approach) has been adopted for this eigenvalue problem. For a fixed energy, the problem is solvable for a set of values of the potential parameters (the "parameter spectrum"). Subsequently, the map that associates the parameter spectrum with the energy is inverted to give the energy spectrum. The bound states wavefunction is written as a convergent series involving products of the ultra-spherical Gegenbauer polynomial in space and a new polynomial in energy, which is a special case of the "dipole polynomial" of the second kind.

## 1. Introduction

Exactly solvable problems in quantum mechanics are still playing an important role since they contain all necessary information about the quantum system under consideration. In addition some of these potentials can either be associated directly with real physical systems or be used to test the validity of perturbative, numerical and semi-classical approximations of physical systems. Exact solvability of a given Hamiltonian with its boundary conditions entails the exact knowledge of all its eigenfunctions and the corresponding energy spectrum. However, since the early days of quantum mechanics the number of exactly solvable problems is very limited [1]. Still, they played an important role in putting the theory on firm grounds and in improving our understanding of many physical phenomena [2]. The traditional method of finding exact solutions consists in reducing the Schrödinger equation to a given generalized hypergeometric equation [3,4] whose solutions can be mapped, most of the time, to classical orthogonal polynomials. Another approach uses the group algebra to construct the solution in the relevant group representation spaces based on the dynamical symmetry of the physical problem [5]. In recent years there have been some efforts in classifying all types of solvable problems based on symmetry considerations. First, the idea of shape invariance played a major role in classifying exactly solvable nonrelativistic quantum problems in distinct classes.

---

[*] Corresponding Author, Email: haidari@mailaps.org, Fax: +966-26221474. Present temporary address: 1300 Midvale Ave. #307, Los Angeles, CA 90024

Second, new methods were used to generate solutions of solvable models such as supersymmetric quantum mechanics, potential algebra, path integration, and point canonical transformations. Solvable potentials can be classified into exactly solvable, conditionally-exactly solvable, or quasi-exactly solvable. Exactly solvable problems are those for which one can determine the whole spectrum analytically for a continuous range of values of the potential parameters. Conditionally-exactly solvable problems are those for which solutions can be generated only for specific values of the potential parameters whereas, quasi-exactly solvable problems are those for which one can determine only part of the spectrum. These developments were carried out over the years by many authors where several classes of these solutions were accounted for and tabulated (see, for example, the references cited in [1]). In those developments, one ends up looking for a representation in which the Hamiltonian has the diagonal structure $H|\psi_n\rangle = E_n|\psi_n\rangle$ exhibiting the eigenvalues, or spectrum of the associated potential, and corresponding eigenfunctions.

In this work, we search for the bound state solution of the one-dimensional wave equation. Hence, we expand the wave function $\psi$ in the space of square integrable discrete basis elements $\{\phi_n\}_{n=0}^{\infty}$. That is, the wavefunction is expandable as $|\psi(x,E)\rangle = \sum_n f_n(E)|\phi_n(x)\rangle$, where $x$ is the whole real line. The basis functions must be compatible with the domain of the Hamiltonian and satisfy the boundary conditions of the problem. However, the main contribution of this work is in relaxing the usual restriction of a diagonal matrix representation of the eigenvalue wave equation, $\langle\phi_n|H|\phi_m\rangle = E_n\delta_{nm}$. We only require that it be tridiagonal and symmetric. That is, the action of the wave operator on the elements of an orthogonal basis is allowed to take the following general form

$$\langle\phi_n|(H-E)|\phi_m\rangle = (a_n - E)\delta_{n,m} + b_n\delta_{n,m-1} + b_{n-1}\delta_{n,m+1}, \quad (1.1)$$

where the coefficients $\{a_n, b_n\}_{n=0}^{\infty}$ are real and, in general, functions of the potential parameters. The reason why we concentrate on the tridiagonality is to take advantage of the available important mathematical results that associate tridiagonal representations with orthogonal polynomials. Moreover, by relaxing the diagonal constraint then the space of representations becomes large enough to accommodate a larger class of solvable potentials. In fact, we have found the following new solvable hyperbolic single wave potential

$$V(x) = V_0 \frac{\tanh(\lambda x) + \gamma}{\cosh^2(\lambda x)}, \quad (1.2)$$

where $V_0$ and $\gamma$ are real potential parameters and $\lambda$ is a length scale that determines the range of the potential. The first term of this potential is completely new and cannot be predicted by any of the classical approaches based on diagonal representations. However, the second part is a special case of the hyperbolic Pöschl-Teller potential that has an exact conventional (diagonal representation) solution. Physically, the most interesting situation is when the parameter $\gamma$ lies between $-1$ and $+1$ in which case the shape of this potential becomes a hyperbolic single wave about the origin as shown in Figure 1. The potential has two extrema located at

$$x_{\pm} = \frac{1}{\lambda}\tanh^{-1}\left[-\tfrac{1}{3}\left(\gamma \pm \sqrt{\gamma^2 + 3}\right)\right]. \quad (1.3)$$

However, if $|\gamma| \geq 1$ then it becomes a potential well if $\gamma V_0 < 0$ or a potential barrier if $\gamma V_0 > 0$. Therefore, in this article we will be dealing with the case $|\gamma| < 1$ which carries a richer structure. The configuration of this potential allows for both resonances and bound states, which is contrary to the case $|\gamma| \geq 1$ where either bound or resonance energy states are allowed. Nonetheless, our approach can still handle the case $|\gamma| \geq 1$ as long as $\gamma V_0 < 0$.

In section 2, we present the theoretical formulation of the problem and explain how the new solvable hyperbolic single wave potential is obtained. In section 3, we discuss the potential parameter spectrum and how the process can be inverted to generate the usual energy spectrum. To enrich our study we use the newly formulated 1D J-matrix method of scattering (details are in our recent publication [6]) to compute the transmission and reflection coefficients for a given set of potential parameters. Thereafter, we present the solution of the three-term recursion relation and associated wavefunction. Finally, in section 4 we present our conclusions. A brief summary of the main findings in this paper was published recently in a letter [7]

## 2. The Hyperbolic Single Wave Potential

The matrix wave equation is obtained by expanding the wave function $|\psi\rangle$ as $\sum_m f_m |\phi_m\rangle$ in $(H-E)|\psi\rangle = 0$ and projecting on the left by $\langle\phi_n|$. Ensuring a tridiagonal matrix representation of the associated wave equation results in a three-term recursion relation for the expansion coefficients $\{f_n\}_{n=0}^{\infty}$. Consequently, the problem translates into finding solutions of the recursion relation for the expansion coefficients of the wave function $\psi$. In most cases this recursion is solved easily and directly by correspondence with those of well known orthogonal polynomials. Therefore, a solution of the problem is obtained once the expansion coefficients are determined. This approach embodies powerful tools in the analysis of solutions of the wave equation by exploiting the intimate connection and interplay between tridiagonal matrices and the theory of orthogonal polynomials. In such analysis, one is at liberty to employ a wide range of well established methods and numerical techniques associated with these settings such as quadrature approximation and continued fractions [8]. Additionally, since tridiagonal matrices have special and favorable treatments in numerical routines (e.g., in computing their eigenvalues and eigenvectors), the accuracy and convergence of numerical computations are also enhanced.

Let us consider the one dimensional time-independent Schrödinger equation for a point particle of mass $m$ in the field of a potential $V(x)$

$$\left[-\frac{\hbar^2}{2m}\frac{d^2}{dx^2} + V(x) - E\right]\psi(x,E) = 0. \qquad (2.1)$$

The physical configuration space coordinate belongs to the interval $x \in [-\infty, +\infty]$. We make a transformation, $y(\lambda x) = \tanh(\lambda x)$ [9], to a "reference" configuration space with coordinate $y \in [-1, +1]$, where $\lambda$ is a positive scale parameter. This transformation takes the wave equation (2.1) into

$$\left[ -\left(1-y^2\right)\frac{d}{dy}\left(1-y^2\right)\frac{d}{dy} + U(y) - \varepsilon \right]\psi(y,\varepsilon) = 0, \tag{2.2}$$

where $\varepsilon = E/E_0$, $U = V/E_0$, $E_0 = (\lambda\hbar)^2/2m$, and the integration measure is defined as $\int_{-\infty}^{+\infty} \ldots dx = \frac{1}{\lambda}\int_{-1}^{+1} \ldots \frac{dy}{1-y^2}$. Our approach is an algebraic one where we study the solution of the problem by constructing infinite dimensional Hermitian matrices for the wave operator. A complete square integrable basis that is compatible with this problem and carries a faithful description of the wavefunction $\psi(y,\varepsilon)$ has the following basis elements [10]

$$\phi_n(y) = A_n(1+y)^\alpha(1-y)^\beta P_n^{(\mu,\nu)}(y), \tag{2.3}$$

where $P_n^{(\mu,\nu)}(y)$ is the Jacobi polynomial of degree $n = 0,1,2,..$ and $A_n$ is a normalization constant. The real parameters $\mu$ and $\nu$ are larger than $-1$ whereas the values of $\alpha$ and $\beta$ depend on the boundary conditions and square integrability of the eigenstates. The matrix elements of a given function $F(x)$ in this basis becomes

$$F_{nm} = \langle\phi_n|F|\phi_m\rangle = \frac{A_n A_m}{\lambda}\int_{-1}^{+1}(1+y)^{2\alpha-1}(1+y)^{2\beta-1}P_n^{(\mu,\nu)}(y)P_n^{(\mu,\nu)}(y)F(x(y))dy \tag{2.4}$$

The scalar product is a special case where $F(x) = 1$. It is only for a limited and special choice of functions and values of the basis parameters that this matrix representation becomes tridiagonal. For example, the scalar product is diagonal if $(\mu,\nu) = (2\beta-1, 2\alpha-1)$ and it becomes tridiagonal if $(\mu,\nu) = (2\beta-1, 2\alpha-2)$ or if $(\mu,\nu) = (2\beta-2, 2\alpha-1)$. If we define the wave operator $J$ as

$$J(y) = H - \varepsilon = -(1-y^2)\tfrac{d}{dy}(1-y^2)\tfrac{d}{dy} + U(y) - \varepsilon, \tag{2.5}$$

then its matrix elements in the basis (2.3) could also become tridiagonal only for a given special set of potential functions $U(y)$. That is, under this requirement $J_{nm} = \langle\phi_n|J|\phi_m\rangle = 0$ for all $|n-m| \geq 2$. This requirement could be easily supported if we keep in mind the orthogonality relation of the Jacobi polynomials, their associated weight function, and recursion relations as shown in the Appendix of paper I [11]. Applying the differential wave operator (2.5) on the basis $|\phi_m\rangle$ then projecting on the left by $\langle\phi_n|$, we obtain the following matrix elements of the wave operator

$$J_{mn} = \langle\phi_m|\left\{(1-y^2)\left[n\left(y + \tfrac{\nu-\mu}{2n+\mu+\nu}\right)\left(\tfrac{2\alpha-\nu}{1+y} - \tfrac{2\beta-\mu}{1-y}\right) + 2\alpha\beta + n(n+\mu+\nu+1) + \alpha + \beta \right.\right.$$
$$\left.\left. -\beta^2\tfrac{1+y}{1-y} - \alpha^2\tfrac{1-y}{1+y}\right] + U(y) - \varepsilon\right\}|\phi_n\rangle - 2\tfrac{A_n(n+\mu)(n+\nu)}{A_{n-1}(2n+\mu+\nu)}\langle\phi_m|\left(\tfrac{2\alpha-\nu}{1+y} - \tfrac{2\beta-\mu}{1-y}\right)|\phi_{n-1}\rangle \tag{2.6}$$

Requiring a tridiagonal representation imposes the following conditions on the last term which should either be eliminated by choosing $(\alpha,\beta) = \left(\tfrac{\nu}{2},\tfrac{\mu}{2}\right)$ or be proportional to $\delta_{m,n-1}$, which requires that $(\alpha,\beta)$ be either $\left(\tfrac{\nu+1}{2},\tfrac{\mu}{2}\right)$ or $\left(\tfrac{\nu}{2},\tfrac{\mu+1}{2}\right)$. Further development shows that the last two cases correspond to the hyperbolic Rosen-Morse potential that has already been treated extensively in the literature [2,12]. Hence, we will only be concerned with the first case that results in the following matrix elements of the wave operator

$$J_{mn} = \langle\phi_m|(1-y^2)\left[n(n+\mu+\nu+1) + \tfrac{1}{2}(\mu\nu+\mu+\nu) - \left(\tfrac{\mu}{2}\right)^2\tfrac{1+y}{1-y} - \left(\tfrac{\nu}{2}\right)^2\tfrac{1-y}{1+y}\right]|\phi_n\rangle$$
$$+ \langle\phi_m|(U-\varepsilon)|\phi_n\rangle \tag{2.7}$$

Evaluation of the above matrix elements, with the help of the properties of the Jacobi polynomials [13], shows that $\langle \phi_m | (1-y^2) | \phi_n \rangle = \delta_{m,n}$ whereas $\langle \phi_m | y(1-y^2) | \phi_n \rangle$ is the only allowed tridiagonal element. All other elements destroy the tridiagonal structure and hence should be eliminated by appropriate counter terms in the potential. Thus, we see from (2.7) that our solvable potential should have the following form

$$U(y) - \varepsilon = (1-y^2)\left[ \left(\tfrac{\mu}{2}\right)^2 \frac{1+y}{1-y} + \left(\tfrac{\nu}{2}\right)^2 \frac{1-y}{1+y} + Cy + D \right] \tag{2.8}$$

In terms of the original configuration space coordinate $x$, we can write the potential as

$$\frac{1}{E_0} V(x) = A \tanh(\lambda x) + \frac{B}{\cosh^2(\lambda x)} + C \frac{\tanh(\lambda x)}{\cosh^2(\lambda x)}, \tag{2.9}$$

where $A$, $B$, and $C$ are real dimensionless potential parameters and $D = B - \tfrac{1}{2}\varepsilon$. The basis parameters are related to the potential parameters and energy as follows

$$\mu^2 - \nu^2 = 2A, \quad \mu^2 + \nu^2 = -2\varepsilon. \tag{2.10}$$

This makes the basis parameters $\mu$ and $\nu$ energy dependent and dictates that the problem is solvable only for bound states ($\varepsilon < 0$). The first and second potential terms in (2.9) constitute the hyperbolic Rosen-Morse potential which is well-known and has been listed among the exactly solvable potentials in the literature [1,14]. On the other hand, the last term is new and constitutes the main component in the new solvable potential. Obviously, diagonalizing the Hamiltonian with this potential will not lead to an exact solution unless $C = 0$. This is the reason why we relax the diagonal constraint by working in a more general tridiagonal representation that makes it possible to search for such a solution, if it existed. To make the new potential richer in its spectra we combine the second and last term in (2.9) to obtain

$$\frac{1}{E_0} V(x) = C \frac{\tanh(\lambda x) + \gamma}{\cosh^2(\lambda x)}, \tag{2.11}$$

where $\gamma = B/C$ for $C \neq 0$ and $V_0 = E_0 C$. The above potential corresponds to the choice $\mu^2 = \nu^2$ (i.e., $A = 0$), which also makes the potential short-range (i.e., $\lim_{x \to \pm\infty} V(x) = 0$). With these results and for the case $\mu = \nu = \sqrt{-\varepsilon}$, the basis functions could be written as

$$\phi_n(x) \sim (\cosh \lambda x)^{-\mu} C_n^{\mu+\frac{1}{2}} (\tanh \lambda x), \tag{2.12}$$

where $C_n^\nu(z)$ is the Gegenbauer ultra-spherical polynomial [13]

## 3. Potential Parameter Spectrum, Energy Spectrum, and Eigenfunctions

Expanding the wavefunction in the complete basis $\{|\phi_n\rangle\}_{n=0}^\infty$ as the infinite sum $|\psi\rangle = \sum_{n=0}^\infty f_n |\phi_n\rangle$ makes the wave equation (2.2) equivalent to the following three term recursion relation for expansion coefficients $\{f_n\}_{n=0}^\infty$

$$J_{nn} f_n + J_{nn-1} f_{n-1} + J_{nn+1} f_{n+1} = 0, \tag{3.1}$$

where the matrix wave operator is given by (with $\mathcal{A}_n = \sqrt{\frac{2n+\mu+\nu+1}{2^{\mu+\nu+1}} \frac{\Gamma(n+1)\Gamma(n+\mu+\nu+1)}{\Gamma(n+\nu+1)\Gamma(n+\mu+1)}}$ )

$$J_{nm} = \left[ \gamma C + \left(n + \tfrac{\nu+\mu}{2}\right)\left(n + \tfrac{\nu+\mu}{2} + 1\right) \right] \delta_{nm} + C \langle n | y | m \rangle, \tag{3.2}$$

and $\langle n | y | m \rangle$ has the following tridiagonal matrix representation [11]

–5–

$$\langle n|y|m\rangle = \frac{v^2-\mu^2}{(2n+\mu+v)(2n+\mu+v+2)}\delta_{n,m}$$
$$+\frac{2}{2n+\mu+v}\sqrt{\frac{n(n+\mu)(n+v)(n+\mu+v)}{(2n+\mu+v-1)(2n+\mu+v+1)}}\delta_{n,m+1} \quad (3.3)$$
$$+\frac{2}{2n+\mu+v+2}\sqrt{\frac{(n+1)(n+\mu+1)(n+v+1)(n+\mu+v+1)}{(2n+\mu+v+1)(2n+\mu+v+3)}}\delta_{n,m-1}$$

The off-diagonal entries in the wave operator matrix (3.2) are due only to the last term, which is proportional to the potential parameter $C$. Recall that our hyperbolic single wave potential (2.11) holds for two cases corresponding to $v=\pm\mu$. The recursion relation is obtained by substituting the matrix elements of the wave operator given by Eq. (3.2) into Eq. (3.1). The result for the case $v=+\mu$ ($>-1$) is as follows

$$-\gamma f_n = C^{-1}a_n f_n + b_{n-1}f_{n-1} + b_n f_{n+1}, \quad (3.4)$$

where $a_n = (n+\mu)(n+\mu+1)$ and $b_n = \frac{1}{2}\sqrt{\frac{(n+1)(n+2\mu+1)}{(n+\mu+1)^2-1/4}}$. This recursion relation is a special case associated with the "dipole polynomials" of the second kind, $G_n^{(\rho,\sigma)}(\kappa;z)$ [15], with $\rho=\sigma=\mu$, $\kappa=-\frac{1}{2}C$, and $z=\gamma C-\frac{1}{4}$. Doing the same as above for the case $v=-\mu$, which requires $-1<\mu<+1$, leads to the following recursion relation for the wavefunction expansion coefficients

$$-\gamma f_n = n(n+1)C^{-1}f_n + \frac{1}{2}\sqrt{\frac{n^2-\mu^2}{n^2-1/4}}f_{n-1} + \frac{1}{2}\sqrt{\frac{(n+1)^2-\mu^2}{(n+1)^2-1/4}}f_{n+1}. \quad (3.5)$$

In both cases one is dealing with a generalized eigenvalue problem since the energy eigenvalue $\varepsilon$ is buried in the basis parameter $\mu(\varepsilon)=\pm\sqrt{-\varepsilon}$, which in its turn appears in all matrix elements. Because the basis (2.3) is energy dependent, through the parameter $\mu(\varepsilon)$, our solution strategy will be different from that in paper I. For a given (negative) value of the energy, we find the set of values of the potential parameters that leads to an exact solution. Depending on the energy and physical constraints, this set could be finite or infinite. We call this set, the "potential parameter spectrum" or simply the *parameter spectrum*. The concept of parameter spectrum was introduced for the first time in the solution of the wave equation in [16]. If the map that associates the parameter spectrum with the energy is invertible, then we could easily obtain the *energy spectrum* for a given choice of potential parameters. Multiplying both sides of Eq. (3.4) by $\sqrt{a_n}$ and defining a new polynomial $g_n$ as $g_n = \sqrt{\frac{a_n}{a_0}}f_n$, we obtain a recursion relation for $g_n$. Dividing both sides of this new relation by $a_n$, we get the following

$$-C^{-1}g_n = A_n g_n + B_n g_{n+1} + B_{n-1} g_{n-1} \quad (3.6)$$
$$-C^{-1}g_0 = A_0 g_0 + B_0 g_1 \quad (3.7)$$

where and $A_n = \gamma/a_n$, $B_n = b_n/\sqrt{a_n a_{n+1}}$. This recursion relation could be cast in the form of an eigenvalue equation, $T_\gamma|g\rangle = -C^{-1}|g\rangle$, where $|g\rangle$ is the eigenvector and $-C^{-1}$ is the eigenvalue. Thus, the elements of the matrix $T_\gamma$ are obtained as follows

$$(T_\gamma)_{n,m} = A_n\delta_{n,m} + B_n\delta_{n,m-1} + B_{n-1}\delta_{n,m+1}. \quad (3.8)$$

For a given negative energy $\varepsilon$ (equivalently, $\mu$) and parameter $\gamma$, this eigenvalue equation gives an infinite set of discrete values for the potential strength (the $C$-parameter spectrum). They correspond to the set of all problems with these potential strengths whose energy spectra contain the chosen energy $\varepsilon$. The new recursion coefficients $A_n$ and $B_n$ approach the limit of large $n$ as $n^{-2}$. Thus, using (3.6) to calculate the $C$-parameter spectrum gives a more rapidly convergent result than using (3.4) to calculate the $\gamma$-parameter spectrum. Figure 2 shows the calculated potential strength for a fixed parameter $\gamma$ and for all bound states in a properly chosen energy range. The figure is shown with $C$ on the horizontal axis and $-\varepsilon$ on the vertical axis to make it more convenient to visualize the energy spectrum. Thus, a vertical line that crosses the $C$-axis at any chosen potential strength value, say $C'$, intersects the curves at the energy spectrum corresponding to the potential with parameters $C'$ and $\gamma$. Out of these values, the most interesting are those at zero energy (i.e., at the boundary of the energy spectrum). We designate this sub-subset by the symbol $\{\hat{C}_n(\gamma)\}$ and list some of these values in the Table for several values of $\gamma$. At these critical values, the state experiences transition from bound to resonance or vice versa (similar phenomenon was observed for the Yukawa potential [17]). It is evident form Figure 2 that for a given potential strength $C$ in the range $\hat{C}_n < C < \hat{C}_{n+1}$, the system will have $n+1$ bound states. Therefore, the set $\{\hat{C}_n(\gamma)\}$ is very important for bound states number counting. It is interesting to note that there is no minimum critical potential strength in this 1D case for $\gamma V_0 > 0$. A similar analysis can be performed for the case $\nu = -\mu$ which leads to a recursion relation similar to (3.6) but with $a_n = n(n+1)$ and $b_n = \frac{1}{2}\sqrt{\frac{(n+1)^2 - \mu^2}{(n+1)^2 - 1/4}}$; then we proceed similarly to the previous analysis keeping in mind that, in this case, $-1 < \mu < +1$. As stated above, solvability of this problem is confined to bound states. However, using the tools of the J-matrix method of scattering in 1D [6] we can obtain a highly accurate evaluation of the reflection and transmission amplitudes, $R(E)$ and $T(E)$, for a given potential parameters as shown in Figure 3.

It is worth noting that the current problem has an interesting symmetry that we label as "$\mathcal{CP}\gamma$ symmetry". It amounts to invariance under the following transformation
$$V_0 \to -V_0, \; x \to -x, \; \gamma \to -\gamma. \tag{3.9}$$
Thus, the energy spectra and wavefunctions are also invariant under this transformation. Now, the solution of the three-term recursion relation (3.4) for a given energy is defined modulo an overall non-singular function of the potential parameters $C$ and $\gamma$. If we call this function $\omega^\mu(\gamma, C)$, then we can write $f_n(\varepsilon) = \omega^\mu(\gamma, C) P_n^\mu(\gamma, C)$ where the energy dependence is carried by the parameter $\mu$. Substituting this in the recursion (3.4) and choosing the standard normalization $P_0^\mu = 1$ determines $P_n^\mu(\gamma, C)$ as polynomials of degree $n$ in $\gamma$ and $C^{-1}$ for all $n$. For example, the first few are
$$P_0^\mu(\gamma, C) = 1 \tag{3.10a}$$
$$P_1^\mu(\gamma, C) = -\sqrt{2\mu+3}\left[\gamma + \mu(\mu+1)C^{-1}\right] \tag{3.10b}$$
$$P_2^\mu(\gamma, C) = \tfrac{2}{2\mu+3}\sqrt{\tfrac{\mu+1}{2\mu+5}}\left[\gamma + \mu(\mu+1)C^{-1}\right]\left[\gamma + (\mu+1)(\mu+2)C^{-1}\right] - \tfrac{1}{2}\sqrt{\tfrac{2\mu+5}{\mu+1}} \tag{3.10c}$$
………..
………..

$$P_n^\mu(\gamma,C) = -\frac{1}{b_{n-1}}\left[\left(\gamma + C^{-1}a_{n-1}\right)P_{n-1}^\mu(\gamma,C) + b_{n-2}P_{n-2}^\mu(\gamma,C)\right] \quad (3.10d)$$

Completeness of the basis and normalization of the wavefunction give $\omega^\mu(\gamma,C)$ as the inverse of the square root of the kernel operator associated with these polynomials at the infinite order limit. For details, see paper I [11] and II [17]. The wavefunction that corresponds to a bound state with energy $\varepsilon_n$ is expressed as $\psi(x,\varepsilon_n) \approx \omega^{\mu_n}(\gamma,C)$ $\times \sum_{m=0}^{N-1} P_m^{\mu_n}(\gamma,C)\phi_m(x)$, for some large enough integer $N$ and where $C$ and $\gamma$ belong to the parameter spectrum associated with $\varepsilon_n$. Numerically, we find that the sum converges quickly but becomes unstable if the number of terms, $N$, becomes too large exceeding an integer that depends on the potential parameters and energy level. Moreover, trying to evaluate the wavefunction at an energy that does not belong to the spectrum will never achieve stable results. It will only produce rapidly increasing oscillations with large amplitudes. In fact, the sum of these oscillations for large $N$ leads to destructive interference that should result in zero net value for the wavefunction. Finally, we write the complete bound state wavefunction as

$$\begin{aligned}\psi(r,\varepsilon_n) = &\,\omega^{\mu_n}(\gamma,C)\pi^{-\frac{1}{2}}2^{\mu_n}\Gamma\left(\mu_n+\tfrac{1}{2}\right)(\cosh\lambda x)^{-\mu_n} \\ &\times \sum_{m=0}^{\infty}\sqrt{\left(m+\mu_n+\tfrac{1}{2}\right)\tfrac{\Gamma(m+1)}{\Gamma(m+2\mu_n+1)}}\,P_m^{\mu_n}(\gamma,C)\mathcal{C}_m^{\mu_n+\frac{1}{2}}(\tanh\lambda x)\end{aligned} \quad (3.11)$$

where $\mu_n = \sqrt{-\varepsilon_n}$.

## 4. Conclusion

By working in a complete square integrable basis that carries a tridiagonal matrix representation for the wave operator, we have succeeded in finding a new solvable hyperbolic single wave potential in one dimension. The notion of *exact solvability* was defined in our present work as the ability to calculate all physical quantities in the problem to any desired accuracy limited only by the computing machine precision; no physical approximations are invoked. The tridiagonal matrix representation for the wave operator is equivalent to a three-term recursion relation for the expansion coefficients of the wave function in the basis. Finding solutions of the recursion relation is equivalent to solving the original problem. However, our solution strategy of the three term recursion relation differs from that in our previous work [11,18] in that the basis is energy dependent. Thus we first obtained the "potential parameter spectrum", defined as being the set of values of the potential parameters that leads to an exact solution of the eigenvalue equation for a given value of the energy. Thereafter, the map that associates the parameter spectrum with the energy is inverted to enable us to obtain the *energy spectrum* very accurately. The bound state wavefunction is also expressed in closed form as a convergent series in terms of orthogonal polynomials. We expect that all other physical quantities can be computed to a high degree of accuracy using our present approach.

The difference between our present tridiagonalization approach and the direct numerical integration approach is that we obtain a closed form three term recursion relation for the wavefunction expansion coefficients. We have also managed to have these expansion coefficients decrease fast enough asymptotically to ensure fast convergence and controllable numerical accuracy. We believe that the tridiagonal representation approach will enable us to enlarge, relatively, the class of exactly solvable quantum problems in all space dimensions and that it could easily be extended to non-central [19] as well as relativistic problems.

**Acknowledgements:** This work is partially sponsored by King Fahd University of Petroleum and Minerals. We also acknowledge the support of the Saudi Center for Theoretical Physics.

**Table caption:** The critical potential strength parameter for several positive values of $\gamma$. For negative values, we obtain $\hat{C}_n(-\gamma) = -\hat{C}_n(\gamma)$.

**Table**

| $n$ | $\hat{C}_n(0.2)$ | $\hat{C}_n(0.4)$ | $\hat{C}_n(0.6)$ | $\hat{C}_n(0.8)$ |
|---|---|---|---|---|
| 0 | 9.4299992413 | 16.1287906278 | 34.2552861086 | 124.1641648307 |
| 1 | 41.7931015925 | 73.8722073011 | 163.6321410556 | 632.3975147612 |
| 2 | 96.5065433233 | 172.6156423881 | 387.9808630087 | 1530.9247509090 |
| 3 | 173.5087786214 | 312.2798396323 | 707.1697277952 | 2819.3834264375 |
| 4 | 272.7870082299 | 492.8478458470 | 1121.1705816654 | 4497.6964255837 |
| 5 | 394.3368379360 | 714.3138793839 | 1629.9739208542 | 6565.8380267476 |
| 0 | 0 | 0 | 0 | 0 |
| 1 | −4.4155383280 | −3.3249120592 | −2.6180242812 | −2.1359006835 |
| 2 | −18.4182760066 | −13.3678268362 | −10.0857158881 | −7.8729202472 |
| 3 | −41.6866325080 | −29.9446503029 | −22.2777418206 | −17.0399291212 |
| 4 | −74.1505686365 | −52.9917122329 | −39.1800751768 | −29.6522394177 |
| 5 | −115.7969855345 | −82.4884631605 | −60.7733768741 | −45.7189890761 |

**Figure captions:**

**Fig. 1:** The potential function (1.2) (in units of $|V_0|$) versus the $x$-coordinate (in units of $\lambda^{-1}$) for several values of $\gamma$ ranging from $-1$ to $+1$.

**Fig. 2:** Energy spectrum associated with the potential (2.11) as a function of the potential strength $V_0 = E_0 C$ and for $\gamma = -\frac{1}{2}$.

**Fig. 3:** Reflection and transmission coefficients as a function of the energy (in units of $V_0$) and for $\gamma = \frac{1}{5}$.

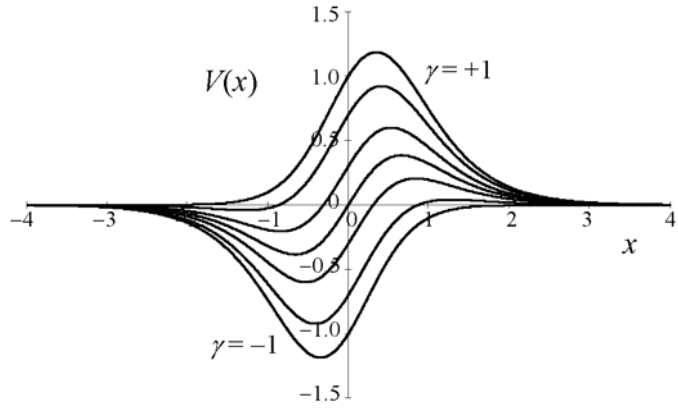

**Fig. 1**

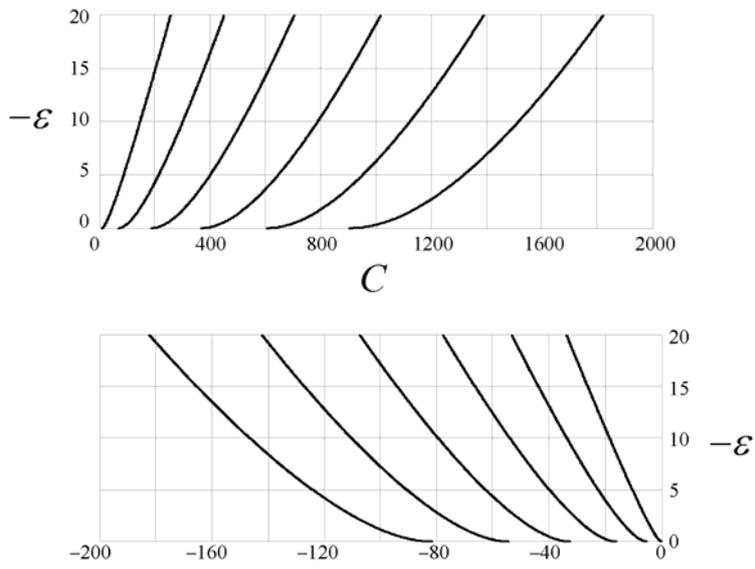

**Fig. 2**

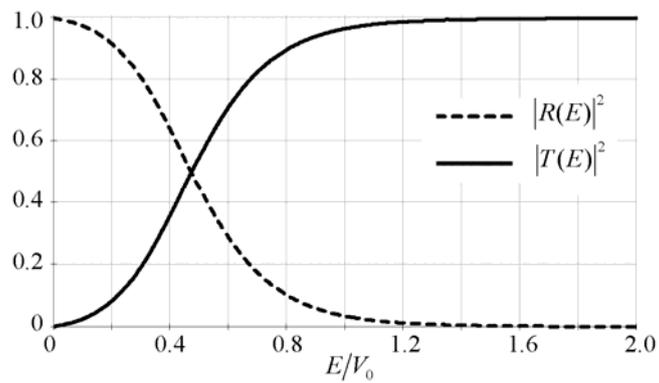

**Fig. 3**